\documentclass{aa}

\usepackage{graphicx}
\usepackage{txfonts}
\begin{document} 

%	- Nome do catalogo (hipoteses):
%		- PlaSt, ExoPlaSt, SP^2 (Spectroscopic Parameters for Stars with Planets), SPEC (Spectroscopic Parameters
%		  for Exoplanet host stars Catalogue), SWEET (catalog of parameters for Star With ExoplanETs)

   \title{SWEET-Cat: A catalogue of parameters for Stars With ExoplanETs\thanks{Based on observations collected at ESO facilities
   under programs 088.C-0892(A), 089.C-0444(A), 090.C-0146(A) (FEROS spectrograph, 2.2-m ESO-MPI telescope, La Silla), 380.C-0083(A), and 083.C-0174(A) (UVES spectrograph, ESO VLT Kueyen telescope, Paranal). }}

  \subtitle{{I. New atmospheric parameters and masses for 48 stars with planets}}

   \author{N.C. Santos\inst{1,2}
          \and
          S. G. Sousa\inst{1,4}
          \and
          A. Mortier\inst{1,2}
          \and 
          V. Neves\inst{1,2,3}
          \and
          V. Adibekyan\inst{1}
          \and
          M. Tsantaki\inst{1,2}
          \and
          E. Delgado Mena\inst{1}
          \and
          X. Bonfils\inst{3}
          \and 
          G. Israelian\inst{4,5}
          \and
          M. Mayor\inst{6}
          \and
          S. Udry\inst{6}
          }

   \institute{     
         Centro de Astrof\'{\i}sica, Universidade do Porto, Rua das Estrelas, 4150-762 Porto, Portugal         
         \and
         Departamento de F\'{\i}sica e Astronomia, Faculdade de Ci\^encias, Universidade do Porto, Rua do Campo Alegre, 4169-007 Porto, Portugal
         \and
         UJF-Grenoble 1 / CNRS-INSU, Institut de Plan\'etologie et d'Astrophysique de Grenoble (IPAG) UMR 5274, Grenoble, F-38041, France
         \and
	Instituto de Astrof{\'\i}sica de Canarias, E-38200 La Laguna, Tenerife, Spain
	\and
	Departamento de Astrof{\'\i}sica, Universidad de La Laguna, E-38206 La Laguna, Tenerife, Spain
         \and
	Observatoire de Gen\`eve, Universit\'e de Gen\`eve, 51 ch. des Maillettes, CH-1290 Sauverny, Switzerland
        }

   \date{Received XXX; accepted XXX}

% \abstract{}{}{}{}{} 
% 5 {} token are mandatory
 
  \abstract
   {Due to the importance that the star-planet relation has to our understanding of the planet formation process, the precise determination of stellar parameters for 
   the ever increasing number of discovered extra-solar planets is of great relevance. Furthermore, precise stellar parameters are needed to
   fully characterize the planet properties. It is thus important to continue the efforts to determine, in the most uniform way possible, the parameters for
   stars with planets as new discoveries are announced. 
   %Unfortunately, the existing data for planet-host stars
   %is usually dispersed in the literature, and difficult to compile. The few compilations that exist are also not complete or do
   %not put their focus on uniformity.
   }
   {In this paper we present new precise atmospheric parameters for a sample of 48 stars with planets.
   We then take the opportunity to present a new catalogue of stellar parameters for FGK and M stars with planets detected by radial velocity, 
   transit, and astrometry programs.}
   {Stellar atmospheric parameters and masses for the 48 stars were derived assuming LTE and using high resolution and high signal-to-noise spectra. 
   The methodology used is based on the measurement of equivalent widths for a list of iron lines and making use of iron ionization and excitation equilibrium
   principles.
   For the catalog, and whenever possible, we used parameters derived in previous works published by our team,
   using well defined methodologies for the derivation of stellar atmospheric parameters. This set of parameters amounts 
   to over 65\% of all planet host stars known, including more than 90\% of all stars with planets discovered through radial velocity surveys. For the remaining targets, 
   stellar parameters were collected from the literature.}
   {The stellar parameters for the 48 stars are presented and compared with previously determined literature values. For the catalog, we compile values for the effective temperature, surface gravity, metallicity, and stellar mass for (almost) all the planet host stars listed in the Extra-solar Planets Encyclopaedia.
   This data will be updated on a continuous basis. The compiled catalogue is available online. The data can be used for statistical studies of the star-planet correlation, as well as for the derivation of consistent properties for known planets.}
  {}
 %  {The compiled catalogue is available online. The data can be used for statistical studies of the star-planet correlation, as well as for the derivation of consistent 
 %properties for known planets.}

   \keywords{planetary systems --
                Stars: solar-type --
                Stars: abundances --
                Catalogs
               }

\maketitle

\section{Introduction}
\label{sec:intro}

The study of extrasolar planetary systems is steadily becoming a mature field of research. To date,
over 850 extra-solar planets have been discovered around solar-type stars\footnote{For an updated table
we point to \url{http://www.exoplanet.eu}}. Most of these were
found thanks to the incredible precision achieved by today's radial velocity and photometric transit techniques.
On top of the dozens of giant planets detected, these efforts are adding to the lists the first planets that may 
be rocky in nature like our Earth \citep[e.g.][]{Leger-2009,Batalha-2011,Dumusque-2012}.
To these we should add a plethora of additional candidates announced as part of space based
transit surveys like Kepler \citep[][]{Batalha-2013}. Overall, these discoveries are showing that planets are 
ubiquitous around solar-type stars \citep[e.g.][]{Mayor-2011,Howard-2012}.

The strong increase in the number of known planetary systems is allowing astronomers
to analyze in a statistically significant way the properties of the newfound worlds \citep[see e.g.][]{Udry-2007}.
In addition, a combination of different techniques and methods is also giving the possibility to explore
the planetary properties, including the study of their atmospheres and internal 
structure \citep[e.g.][]{Valencia-2010,Cowan-2011,Demory-2012}. 

A key aspect in all this progress is the characterization of the planet host stars. Several reasons exist for that.
For instance, precise (or if possible, accurate) stellar radii are critical if we want to measure precise values
for the radius of a transiting planet \citep[see e.g.][]{Torres-2012}. The determination of stellar radii depends, on its hand, 
on the quality of the derived stellar parameters such as the effective temperature. 

The chemical composition 
of a planet, both its interior and atmosphere, is also likely to be related to the chemical composition of
the proto-stellar cloud, reflected on the composition of the stellar atmosphere \citep[][]{Guillot-2006,Fortney-2007,Bond-2010}.
The precise derivation of stellar chemical abundances thus gives us important clues to understand
the planets and their observed properties. %and their formation process. %Finally, the amount of stellar irradiation, depden may also play a critical role in the radii of giant planets \citep[e.g.][]{Guillot-2002}.

Further to this, a number of studies have pointed towards the existence of a strong relation between the properties 
and frequency of the newfound planets and those of their host stars. 
In this respect, the well known correlation between the stellar metallicity and the frequency of giant planets
is a good example. Large spectroscopic studies \citep[e.g.][]{Santos-2001,Santos-2004b,Fischer-2005,Sousa-2011,Mayor-2011,Mortier-2013a}
confirmed the initial suspicions \citep[][]{Gonzalez-1997,Santos-2000b} of a positive correlation between
the probability of finding a giant planet and the metal content of the stars. This strong correlation even prompted
new planet search surveys based on metal-rich samples \citep[e.g.][]{Tinney-2003,Fischer-2004,DaSilva-2006}.
Although positively increasing the planet detection rate, these surveys biased the samples towards metal-rich
stars, a bias that has to be taken into account when studying the metallicity-planet correlation.

Curiously, this strong metallicity-giant planet correlation was not
found for the lowest mass planets \citep[][]{Sousa-2011,Mayor-2011,Buchhave-2012}. Both results, however,
are in full agreement with the expectations from the most recent models of planet 
formation based on the core-accretion paradigm \citep[e.g.][and discussion therein]{Mordasini-2012}.

Although the general metallicity-giant planet correlation is reasonably well established, many details are
still missing that may hold the clue to new and important details concerning planet formation. For example,
the exact shape of the metallicity-planet correlation is still debated \citep[][]{Santos-2004b,Johnson-2010,Mortier-2013a}.
The understanding of this issue may be critical to point out the mechanisms responsible for the
formation of giant planets across the whole metallicity range \citep[e.g.][]{Matsuo-2007}, or to
the understanding of the frequency of planets in the Milky Way. %\citep[see e.g.][]{Lineweaver-2004}.
The role of the abundances of other elements is also being discussed \citep[e.g.][]{Adibekyan-2012}, with some 
curious trends being a strong matter of debate, concerning e.g. the abundances of the light element 
lithium \citep[][]{Israelian-2009,Baumann-2010,Sousa-2010b,Ghezzi-2010b}
or specific trends including other elemental abundances \citep[e.g.][]{Ramirez-2010,Gonzalez-Hernandez-2010}.

Similar to the stellar metallicity, stellar mass has also been pointed out to play a role in the
formation of giant planets. It is now widely accepted that the frequency of giant planets orbiting (lower mass)
M-dwarfs is considerably lower than the one found
for FGK dwarfs \citep[][]{Bonfils-2005b,Bonfils-2011,Endl-2006}, at least regarding the short period domain \citep[][]{Neves-2013}.
Higher mass stars, on the other hand, seem to have a higher frequency
of orbiting giant planets \citep[][]{Lovis-2007,Johnson-2007}. This result is expected from the models of planetary
formation following the core-accretion paradigm \citep[][]{Laughlin-2004,Ida-2005,Kennedy-2008} -- see however \citet[][]{Kornet-2005,Boss-2006}.
Note that this correlation may be related to the different trend in stellar metallicity that has been suggested
to exist for intermediate mass giant stars with planets \citep[][]{Pasquini-2007,Ghezzi-2010,Hekker-2007}.

Finally, it is important to note that the role of stellar properties (metallicity, temperature) on the formation of different architectures of planetary systems has
also been addressed. Among these, suspicions have been raised concerning the metallicity-orbital period relation {\citep[e.g.][]{Queloz-2000,Sozzetti-2004,Santos-2003,Beauge-2013,Dawson-2013},}
with hot-jupiters being often pointed out as orbiting particularly metal-rich stars (note however that this trend has not been confirmed from a statistical point of view). More recently, the temperature and age of the star was shown present a correlation
with the alignement of the stellar spin-orbital plane angle \citep[][]{Winn-2010,Triaud-2011,Albrecht-2012}, a result that hints at the mechanisms responsible for the migration
of hot jupiters.

Paramount to the discussion of all these issues is the correct determination of stellar parameters like the effective temperature, the
stellar metallicity, and the stellar mass. Since accurate values for these are usually not possible\footnote{Possible but debatable exceptions for accurate effective temperature determinations 
may be solar-type dwarfs with accurate parallaxes and interferometric or asteroseismic radii.}, it is critical that at least uniform sets of stellar parameters exist.
Unfortunately this is not always the case, with different teams making use of different methods (line-lists, model atmospheres, methodologies)
to derive the atmospheric properties of the host stars. In many cases, comparisons have shown that the differences are residual \citep[see e.g.][]{Sousa-2008}, but
in other cases the discrepancies have significant impact on the knowledge of the planet parameters \citep[for a recent discussion on the possible offsets see][]{Torres-2012}. 

{
In this paper we present new atmospheric parameters and masses for a sample of 48 stars with planets. The atmospheric parameters were derived in LTE from
a uniform analysis, and making use of high resolution and high S/N spectra. These values are then included in a new
catalog of stellar parameters for stars with planets (that we name SWEET-Cat), also presented in this paper. 
The catalogue, available online, represents an effort to compile a set of data that is usually spread in the literature. The baseline 
parameters in the catalog are also compared with the ones listed in other compilations or catalogs. This comparison provides to
the reader (in particular the exoplanet community) the possibility to understand the typical errors (including systematic) that
exist in the values of parameters for stars with planets published in the literature. 

In the next sections we present the sample of 48 stars discussed in this paper and their stellar parameters. We then present the 
content of the catalogue, the different sources of stellar parameters used, and some considerations about future improvements.}

%	- Comparison of different samples in the literature (a discussion)
%		- Different methods being used: some depend on choice of logg, for example (Torres paper)
%		  e explicar porque
%		- Kepler problem...
%	- Parameters for catalog stars: data, analysis method, comparison 
%		- Mass determinations
%		- Plot of comparison of parameters for different samples: WASP, HAT, Kepler, ...
%	- Stars from the literature
%	- Conclusions

\section{New parameters for 48 planet hosts}
\label{sec:new}

\begin{table*}
\caption{Stellar atmospheric parameters and masses for the 48 planet hosts as presented in this paper.}
\label{TabParNew}
\centering
\begin{tabular}{lccccccl}
\hline\hline
Name & T$_{eff}$ & $\log g_{spec}$ & $\xi$ & [Fe/H] & M$_{\ast}$ & Spectrograph & S/N \\
 & (K) & (dex) & (km s$^{-1}$) & (dex) & (M$_{\odot}$) & & \\
\hline
$\alpha$\,Cen\,B & 5234 $\pm$ 63 & 4.40 $\pm$ 0.11 & 0.90 $\pm$ 0.12 & 0.16 $\pm$ 0.04 & 0.87 $\pm$ 0.07 & HARPS & 1600 \\
BD+144559 & 4864 $\pm$ 101 & 4.26 $\pm$ 0.29 & 0.72 $\pm$ 0.25 & 0.17 $\pm$ 0.06 & 0.82 $\pm$ 0.11 & FEROS & 81 \\
HD7924 & 5133 $\pm$ 68 & 4.46 $\pm$ 0.12 & 0.71 $\pm$ 0.13 & -0.22 $\pm$ 0.04 & 0.77 $\pm$ 0.06 & SOPHIE & 121 \\
HD9578 & 6070 $\pm$ 22 & 4.53 $\pm$ 0.02 & 1.10 $\pm$ 0.03 & 0.16 $\pm$ 0.01 & 1.09 $\pm$ 0.07 & HARPS & 152 \\
HD11506 & 6204 $\pm$ 27 & 4.44 $\pm$ 0.06 & 1.32 $\pm$ 0.03 & 0.36 $\pm$ 0.02 & 1.24 $\pm$ 0.08 & UVES & 114 \\
HD13931 & 5940 $\pm$ 31 & 4.42 $\pm$ 0.07 & 1.19 $\pm$ 0.04 & 0.08 $\pm$ 0.02 & 1.05 $\pm$ 0.08 & SOPHIE & 124 \\
HD16175 & 6030 $\pm$ 22 & 4.23 $\pm$ 0.04 & 1.39 $\pm$ 0.02 & 0.32 $\pm$ 0.02 & 1.26 $\pm$ 0.08 & FIES & 139 \\
HD23127 & 5891 $\pm$ 33 & 4.23 $\pm$ 0.05 & 1.26 $\pm$ 0.04 & 0.41 $\pm$ 0.03 & 1.24 $\pm$ 0.09 & UVES & 81 \\
HD24040 & 5840 $\pm$ 18 & 4.30 $\pm$ 0.03 & 1.14 $\pm$ 0.02 & 0.20 $\pm$ 0.01 & 1.10 $\pm$ 0.08 & UVES & 144 \\
HD27631 & 5700 $\pm$ 20 & 4.37 $\pm$ 0.05 & 1.00 $\pm$ 0.03 & -0.11 $\pm$ 0.02 & 0.94 $\pm$ 0.07 & FEROS & 164 \\
HD31253 & 6147 $\pm$ 22 & 4.27 $\pm$ 0.05 & 1.47 $\pm$ 0.03 & 0.17 $\pm$ 0.02 & 1.23 $\pm$ 0.08 & FEROS & 242 \\
HD33283 & 6058 $\pm$ 30 & 4.16 $\pm$ 0.05 & 1.41 $\pm$ 0.03 & 0.34 $\pm$ 0.02 & 1.33 $\pm$ 0.09 & UVES & 96 \\
HD38283 & 5980 $\pm$ 24 & 4.27 $\pm$ 0.03 & 1.28 $\pm$ 0.03 & -0.14 $\pm$ 0.02 & 1.05 $\pm$ 0.07 & FEROS & 221 \\
HD60532 & 6273 $\pm$ 37 & 4.02 $\pm$ 0.04 & 1.88 $\pm$ 0.05 & -0.09 $\pm$ 0.02 & 1.35 $\pm$ 0.09 & HARPS & 328 \\
HD70573 & 5767 $\pm$ 122 & 4.81 $\pm$ 0.28 & 1.10 $\pm$ 0.26 & -0.18 $\pm$ 0.09 & 0.90 $\pm$ 0.08 & UVES & 160 \\
HD75898 & 6137 $\pm$ 29 & 4.31 $\pm$ 0.05 & 1.36 $\pm$ 0.03 & 0.30 $\pm$ 0.02 & 1.25 $\pm$ 0.08 & UVES & 107 \\
HD77338 & 5440 $\pm$ 52 & 4.36 $\pm$ 0.11 & 1.16 $\pm$ 0.08 & 0.28 $\pm$ 0.04 & 0.97 $\pm$ 0.08 & FEROS & 105 \\
HD86081 & 6036 $\pm$ 23 & 4.21 $\pm$ 0.04 & 1.34 $\pm$ 0.03 & 0.22 $\pm$ 0.02 & 1.23 $\pm$ 0.08 & UVES & 115 \\
HD86226 & 5947 $\pm$ 21 & 4.54 $\pm$ 0.04 & 1.12 $\pm$ 0.03 & 0.02 $\pm$ 0.02 & 1.00 $\pm$ 0.07 & FEROS & 191 \\
HD86264 & 6596 $\pm$ 78 & 4.47 $\pm$ 0.15 & 1.90 $\pm$ 0.09 & 0.37 $\pm$ 0.06 & 1.42 $\pm$ 0.11 & FEROS & 111 \\
HD96167 & 5823 $\pm$ 32 & 4.16 $\pm$ 0.08 & 1.28 $\pm$ 0.03 & 0.38 $\pm$ 0.02 & 1.24 $\pm$ 0.09 & FEROS & 127 \\
HD98649 & 5714 $\pm$ 22 & 4.37 $\pm$ 0.05 & 1.01 $\pm$ 0.03 & -0.03 $\pm$ 0.02 & 0.96 $\pm$ 0.07 & FEROS & 147 \\
HD99109 & 5327 $\pm$ 61 & 4.38 $\pm$ 0.12 & 0.98 $\pm$ 0.09 & 0.30 $\pm$ 0.04 & 0.93 $\pm$ 0.08 & UVES & 59 \\
HD103774 & 6732 $\pm$ 56 & 4.81 $\pm$ 0.06 & 2.03 $\pm$ 0.08 & 0.29 $\pm$ 0.03 & 1.35 $\pm$ 0.09 & HARPS & 257 \\
HD106515A & 5380 $\pm$ 31 & 4.37 $\pm$ 0.05 & 0.81 $\pm$ 0.05 & 0.03 $\pm$ 0.02 & 0.88 $\pm$ 0.06 & HARPS & 129 \\
HD118203 & 5910 $\pm$ 35 & 4.18 $\pm$ 0.07 & 1.34 $\pm$ 0.04 & 0.25 $\pm$ 0.03 & 1.21 $\pm$ 0.09 & SARG & 55 \\
HD126614 & 5601 $\pm$ 44 & 4.25 $\pm$ 0.08 & 1.17 $\pm$ 0.07 & 0.50 $\pm$ 0.04 & 1.14 $\pm$ 0.09 & UVES & 50 \\
HD129445 & 5646 $\pm$ 42 & 4.28 $\pm$ 0.10 & 1.14 $\pm$ 0.06 & 0.37 $\pm$ 0.03 & 1.09 $\pm$ 0.09 & FEROS & 112 \\
HD143361 & 5503 $\pm$ 36 & 4.36 $\pm$ 0.06 & 0.90 $\pm$ 0.06 & 0.22 $\pm$ 0.03 & 0.97 $\pm$ 0.07 & UVES & 73 \\
HD152079 & 5785 $\pm$ 28 & 4.38 $\pm$ 0.05 & 1.09 $\pm$ 0.03 & 0.29 $\pm$ 0.02 & 1.08 $\pm$ 0.08 & HARPS & 115 \\
HD154672 & 5743 $\pm$ 23 & 4.27 $\pm$ 0.04 & 1.08 $\pm$ 0.03 & 0.25 $\pm$ 0.02 & 1.09 $\pm$ 0.08 & UVES & 90 \\
HD155358 & 5908 $\pm$ 28 & 4.26 $\pm$ 0.03 & 1.29 $\pm$ 0.05 & -0.62 $\pm$ 0.02 & 0.91 $\pm$ 0.06 & UVES & 144 \\
HD164509 & 5957 $\pm$ 22 & 4.43 $\pm$ 0.04 & 1.09 $\pm$ 0.03 & 0.24 $\pm$ 0.02 & 1.10 $\pm$ 0.08 & HARPS & 161 \\
HD164604 & 4684 $\pm$ 157 & 4.32 $\pm$ 0.41 & 0.84 $\pm$ 0.33 & 0.12 $\pm$ 0.07 & 0.77 $\pm$ 0.14 & FEROS & 76 \\
HD164922 & 5356 $\pm$ 45 & 4.34 $\pm$ 0.08 & 0.76 $\pm$ 0.08 & 0.14 $\pm$ 0.03 & 0.91 $\pm$ 0.07 & UVES & 86 \\
HD170469 & 5845 $\pm$ 30 & 4.28 $\pm$ 0.13 & 1.17 $\pm$ 0.04 & 0.30 $\pm$ 0.02 & 1.15 $\pm$ 0.09 & UVES & 75 \\
HD175167 & 5635 $\pm$ 28 & 4.09 $\pm$ 0.09 & 1.18 $\pm$ 0.03 & 0.28 $\pm$ 0.02 & 1.17 $\pm$ 0.09 & FEROS & 164 \\
HD176051 & 6030 $\pm$ 41 & 4.68 $\pm$ 0.04 & 1.28 $\pm$ 0.06 & -0.04 $\pm$ 0.03 & 0.99 $\pm$ 0.07 & SOPHIE & 155 \\
HD187085 & 6146 $\pm$ 22 & 4.36 $\pm$ 0.03 & 1.31 $\pm$ 0.03 & 0.13 $\pm$ 0.02 & 1.16 $\pm$ 0.08 & UVES & 169 \\
HD196067 & 5999 $\pm$ 34 & 4.13 $\pm$ 0.04 & 1.30 $\pm$ 0.03 & 0.23 $\pm$ 0.02 & 1.28 $\pm$ 0.09 & FEROS & 113 \\
HD205739 & 6301 $\pm$ 25 & 4.40 $\pm$ 0.04 & 1.42 $\pm$ 0.03 & 0.21 $\pm$ 0.02 & 1.24 $\pm$ 0.08 & UVES & 223 \\
HD207832 & 5736 $\pm$ 27 & 4.51 $\pm$ 0.07 & 1.06 $\pm$ 0.04 & 0.14 $\pm$ 0.02 & 0.98 $\pm$ 0.07 & FEROS & 196 \\
HD218566 & 4808 $\pm$ 85 & 4.09 $\pm$ 0.28 & 0.82 $\pm$ 0.15 & 0.17 $\pm$ 0.04 & 0.86 $\pm$ 0.13 & FEROS & 85 \\
HD220689 & 5904 $\pm$ 26 & 4.38 $\pm$ 0.05 & 1.13 $\pm$ 0.03 & -0.01 $\pm$ 0.02 & 1.02 $\pm$ 0.07 & FEROS & 155 \\
HD220773 & 5995 $\pm$ 34 & 4.26 $\pm$ 0.07 & 1.33 $\pm$ 0.04 & 0.11 $\pm$ 0.03 & 1.15 $\pm$ 0.08 & FEROS & 163 \\
HD224693 & 6053 $\pm$ 28 & 4.18 $\pm$ 0.06 & 1.40 $\pm$ 0.03 & 0.28 $\pm$ 0.02 & 1.29 $\pm$ 0.09 & UVES & 113 \\
HD231701 & 6224 $\pm$ 27 & 4.37 $\pm$ 0.03 & 1.35 $\pm$ 0.03 & 0.04 $\pm$ 0.02 & 1.15 $\pm$ 0.08 & UVES & 145 \\
HIP57274 & 4510 $\pm$ 136 & 4.11 $\pm$ 0.46 & 0.32 $\pm$ 0.59 & 0.01 $\pm$ 0.06 & 0.77 $\pm$ 0.19 & FIES & 70 \\
\hline
\end{tabular}
\end{table*}

{
The sample of 48 stars consists of dwarfs of spectral type F, G, or K that are known to be orbited by a planet found by the radial velocity method (according to the online catalogue \url{www.exoplanet.eu}). The list of stars is presented in Table\,\ref{TabParNew}.

As mentioned above, the parameters were derived from the analysis of high resolution and high S/N spectra. The spectra were gathered through observations, made by our team, and by the use of the ESO archive. In total, six different spectrographs were used: FEROS (2.2m ESO/MPI telescope, La Silla, Chile), FIES (Nordic Optical Telescope, La Palma, Spain), HARPS (3.6m ESO telescope, La Silla, Chile), SARG (TNG Telescope, La Palma, Spain), SOPHIE (1.93m telescope, OHP, France), and UVES (VLT Kueyen telescope, Paranal, Chile). The characreristics of each spectrograph and the number of stars observed are listed in Table\,\ref{TabINS}.

Note that the use of different spectrographs is not expected to introduce significant systematic differences in the derived stellar parameters, as can
be seen from previous studies \citep[e.g.][]{Santos-2004b}.
}

{
The spectra were reduced and extracted using the available pipelines or IRAF \footnote{IRAF is distributed by National Optical Astronomy Observatories, operated by the Association of Universities for Research in Astronomy, Inc., under contract with the National Science Foundation, USA.}. The spectra were then corrected for radial velocity with the IRAF task \texttt{DOPCOR}. Individual exposures of multiple observed stars with the same instrument, were co-added using the task \texttt{SCOMBINE} in IRAF.

From the spectra, we derived the atmospheric stellar parameters (effective temperature T$_{eff}$, surface gravity $\log g$, microturbulence $\xi$ and metallicity [Fe/H]) and the masses as described in Section\,\ref{sec:FGK_RV}. The followed procedure is based on the equivalent widths of \ion{Fe}{i} and \ion{Fe}{ii} lines, and iron excitation and ionization equilibrium, assumed in Local Thermodynamic Equilibrium (LTE). Herefore, the 2002 version of MOOG\footnote{\url{http://www.as.utexas.edu/~chris/moog.html}} \citep{Sneden-1973}, a grid of ATLAS plane-parallel model atmospheres \citep{Kurucz-1993} and the iron linelist of \citet{Sousa-2008} are used. For stars cooler than 5200\,K (as initially derived with the Sousa et al. line list) we re-derived and adopted the parameters using the line list of \citet{Tsantaki-2013}, specially suitable for cool stars. Stellar masses and their errors were computed with the corrected calibration of \citet{Torres-2010} as discussed in Section \ref{sec:FGK_RV}.

To measure the equivalent widths of the iron lines, the code ARES is used \citep[Automatic Routine for line Equivalent widths in stellar Spectra -][]{Sousa-2007}. The input parameters for ARES, are the same as in \citet{Sousa-2008}, except for the \emph{rejt} parameter, which determines the calibration of the continuum position. Since this parameter strongly (and mostly) depends on the S/N of the spectra, different values are needed for each spectrum. In this study, the S/N values were derived for each spectrum using the IRAF routine \texttt{BPLOT}. Three spectral regions are used: [5744\AA, 5747\AA], [6047\AA, 6053\AA] and [6068\AA, 6076\AA]. 
The final S/N of the spectra as measured in the region around 6000\,\AA\ is in most cases above 100 (see Table\,\ref{TabParNew}). Note also that these values
are likely underestimated, since the identification of regions completely absent of absorption lines is not straightforward\footnote{For HARPS spectra, we have in the header an indication about the S/N in each echelle order estimated from the observed flux, but that is not the case for the remaining spectrographs.}.

The \emph{rejt} parameter was then set by visual inspection for 10 different spectra with different S/N values (representable for the whole sample). The \emph{ rejt} parameters for the remaining spectra were then derived by a simple interpolation of these values. This method ensures a uniform usage of the \emph{ rejt} parameter, since we otherwise do not have access to a uniform source for the S/N as in \citet{Sousa-2008}. The final dependence of the \emph{ rejt} parameter to the S/N is the same as in \citet{Mortier-2013b}.
}

\begin{table}
\caption{Spectrograph details: resolving power and spectral ranges.}
\label{TabINS}
\centering
\begin{tabular}{cccc}
\hline\hline
Instrument & Resolving power & Spectral range & Number of \\
 & $\lambda/\Delta\lambda$ & \AA & stars \\
\hline
FEROS & 48000 & 3600 - 9200 & 17 \\
FIES & 67000 & 3700 - 7300 & 2 \\
HARPS & 100000 & 3800 - 7000 & 7 \\
SARG & 57000 - 86000 & 5100 - 10100 & 1 \\
SOPHIE & 75000 & 3820 - 6920 & 3 \\
UVES & 110000 & 3000 - 6800 & 18 \\
\hline
\end{tabular}
\end{table}

{
Table \ref{TabParNew} lists the derived stellar atmospheric parameters for the 48 planet hosts. These dwarf stars cover a wide range in effective temperature, surface gravity, microturbulence, metallicity, and mass: 4510 -- 6732\,K, 4.02 -- 4.81\,dex, 0.32 -- 2.03\,km\,s$^{-1}$, $-$0.62 - 0.5\,dex, and 0.77 -- 1.42\,M$_{\odot}$, respectively. Mean error bars of 42.7\,K, 0.09\,dex, 0.07\,km\,s$^{-1}$, 0.03\,dex, and 0.08\,M$_{\odot}$ are obtained. The errors on the atmospheric parameters were
derived as in \citet[][]{Santos-2004b}. Errors in the mass were computed as described in Sect.\,\ref{sec:FGK_RV}. The uniformity of our analysis minimizes possible systematic errors in the final parameters.

{In Figs.\,\ref{fig:compeu}, \ref{fig:comporg}, and \ref{fig:comparch} we compare our {baseline parameters} for our 48 stars (green symbols) with those listed in the Extrasolar Planets Encyclopaedia \citep[][]{Schneider-2011}, exoplanets.org \citep[][]{Wright-2011}, and the NASA 
Exoplanet Archive\footnote{http://exoplanetarchive.ipac.caltech.edu/}.}
As can be seen from the plots, in general the parameters agree well with previously published values. A few outliers exist, however, in particular
concerning the stellar metallicity (up to $\sim$0.3\,dex). 
}

\section{The SWEET catalogue}
\label{sec:sample}

{As mentioned above, the parameters derived in this paper were added to other values in the literature
into a new catalog of stellar parameters for stars with planets. This catalog is presented in this paper.

The complete list of the fields in the catalogue is listed in Table\,\ref{tab:catalog}. A more detailed description of each field
is given in the following sub-sections.}

\begin{table*}
\caption{List of fields in the catalogue and description (when judged necessary).}
\centering
\begin{tabular}{ll}
\hline\hline
Field & Description\\
\hline
Name & Star name as in the Extra-Solar Planets Encyclopaedia\\
HD number & HD name of the star if available\\
Right Ascension & --\\
Declination & --\\
V magnitude & --\\
Error on V magnitude & --\\
Parallax (mas) & --\\
Error on Parallax (mas) & --\\
Parallax Flag & Source of the parallax measurements\\
Effective temperature (K) & --\\
Error on Effective temperature (K)& --\\
Surface gravity (c.g.s.)&-- \\
Error on Surface gravity (c.g.s.)& --\\
LC Surface gravity (c.g.s.)& Survace gravity from transit light curve \\
Error on LC Surface gravity (c.g.s.)& --\\
Microturbulence (km\,s$^{-1}$) & --\\
Error on microturbulence (km\,s$^{-1}$)& --\\
Metallicity [Fe/H] & --\\    
Error on the metallicity [Fe/H] & --\\    
Stellar mass (M$_\odot$) & --\\
Error on stellar mass (M$_\odot$) & --\\
Sources of parameters with link to ADS& The references are given in the online table \\
Parameter source flag & "1" {for parameters derived by our team (dubbed ``baseline parameters'')}, "0" otherwise\\
Last Update & --\\
Comments & -- \\
%Comments & Any special comments will be included in this column\\
%                     & Whenever available, surface gravities ($\log{g}$) values derived from the analysis
%                     & transit light curve will be included here.
\hline
\end{tabular}
\label{tab:catalog}
\end{table*}

\subsection{Identification and basic data}

At the time that this paper is being written, the Extra-Solar Planets Encyclopaedia lists 889 planets in 694
planetary systems\footnote{As of June 2013}, most of them discovered by radial velocity or transit surveys. Due to its completeness and tradition, we decided to
use this database as a starting point for the catalogue.

For each planet host star listed in the Encyclopaedia as being detected by radial velocity, astrometry, or transit measurements, we compiled a series of basic information.
In this first version of the catalogue we decided to exclude direct imaging planets (most of them around early type stars), planets
discovered using the microlensing technique (due to the difficulty in characterizing the host stars), as well as
degenerated stars (e.g. pulsars hosting planetary systems detected by timing techniques). For the remaining stars (i.e. those listed in the Encyclopaedia as radial velocity, transiting or astrometry planet hosts), we compiled the following basic information.
\begin{itemize}

\item Name of the star: although we adopted the Encyclopaedia name, for all cases where the
star has an HD number, this is also listed;

\item Coordinates: right ascensions and declinations were compiled from Simbad. We adopted the
ICRS coordinates (J2000). Coordinates that were not available were left blank;

\item V magnitudes for all targets were also compiled from Simbad whenever possible.
Exceptions where no V magnitudes were listed in Simbad (for some Kepler and WASP candidates mostly) were taken directly from 
the Encyclopaedia. The V magnitudes are meant to serve as reference, and not to be used for accurate physical calculations;

\item Parallax values were compiled from Simbad whenever these exist. For cases where the
parallax is not available, we computed a ``spectroscopic parallax'' using the estimated stellar parameters
\citep[for details on the method we point to ][]{Sousa-2011b}. For M-dwarfs, a few parallaxes were also taken 
from the literature.

\end{itemize}

\subsection{Atmospheric parameters and masses}
\label{sec:atmospheres}

The determination of accurate stellar atmospheric parameters is a huge matter of debate. Several
methodologies have been explored to derive effective temperatures, surface gravities,
or stellar metallicities, with clear differences in resulting zero point or scale \citep[for some references and discussions see][]{Sousa-2008,Casagrande-2010,Torres-2012}. 
In other words, it is at present very difficult to point towards any accurate source of stellar parameters.
As a consequence, any catalogue of stellar parameters for stars with planets should probably focus, {whenever possible}, 
on uniformity, i.e. on precision rather than accuracy.

With this in mind, in the next sections we describe our baseline sources for the stellar parameters. 
Note however that the catalogue is updated on a regular basis, 
and the sources of parameters may change over time. The philosophy behind the catalogue will
however be maintained.

{The parameters compiled from these {baseline} sources are derived using
a homogeneous analysis {(i.e., as homogeneous as possible, meaning that they were derived by our team using the best possible uniform methodology)}.
We will dub these "baseline parameters" for the rest of the paper, in
opposition with parameters compiled from other literature sources. All together, at the present time, we have
baseline parameters for more than 65\% of all planet host stars, including 87\% of all radial velocity survey planet hosts
(over 95\% if we just include dwarf stars).}

{
We should note that these baseline parameters are the closest that we can have to an homogeneous set
of data. However, the term ``homogeneous'' should be read with some caution. Parameters derived for
different stars, using different data sets (from different spectrographs), cannot be seen as
``fully homogeneous''. For instance, although in most of the cases we use the same baseline line-list to
derive stellar parameters (see below), the final line-list is always a ``sub-sample" of this. This can be due to the fact 
that some spectrographs have spectral gaps, or simply due to the exclusion of some
specific lines, in a case by case analysis, due to the presence of cosmic rays. Adding to this,
the continuum position used when measuring line equivalent widths is subject to errors,
that depend e.g. on the S/N of the data. 
These facts will produce systematic offsets between the parameters derived for the different stars. However,
in the large majority of the cases these offsets are expected to be very small and within the error
bars of the individual parameters. For one example we point the reader to \citet[][]{Sousa-2008} where a
comparison with the parameters derived for a common set of stars using two line-lists (one which
is a sub-sample of the other) is presented. Finally, we cannot exclude systematic effects when comparing the
analysis for stars of significant different temperature or evolutionary stage \citep[e.g.][]{Santos-2009}. These effects are difficult to quantify but 
can be at least partially solved when using appropriate line-lists and
methodologies \citep[see e.g.][]{Tsantaki-2013,Mortier-2013b}.
%In other words, the term ``homogeneous'' should be read ``as homogenous as possible''.
}

\begin{figure}[t!]
\resizebox{\hsize}{!}{\includegraphics{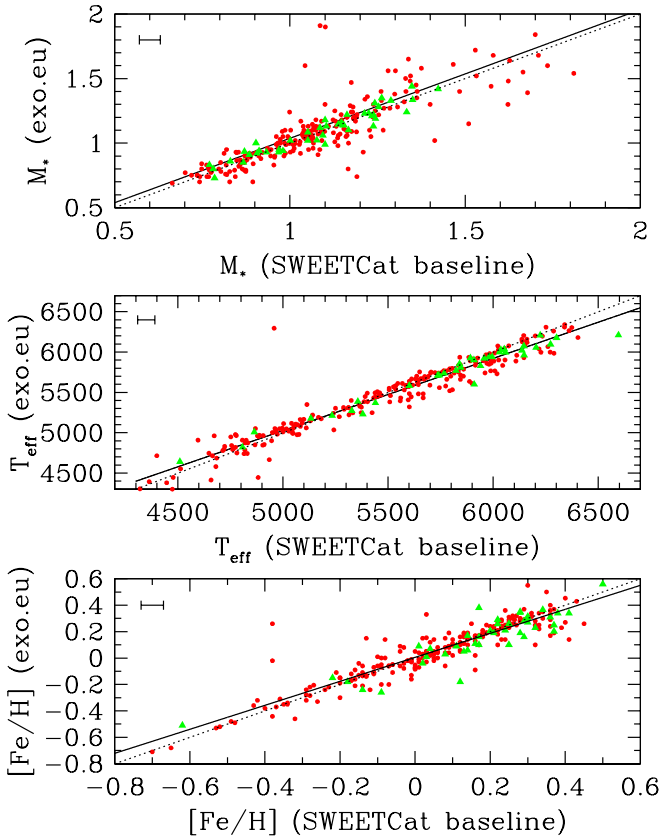}}
\caption{Comparison between our baseline stellar parameters
with those listed in the Extrasolar Planets Encyclopaedia for the same stars. Only stars with planets detected by radial velocity surveys are included. 
{Green triangles denote the 48 stars whose parameters are presented in this paper.}
The dotted line represents a 1:1 relation, and the full line a linear fit to the data. Typical error bars are shown on the upper left part of each panel.}
\label{fig:compeu}
\end{figure}

\begin{figure}[t!]
\resizebox{\hsize}{!}{\includegraphics[angle=0]{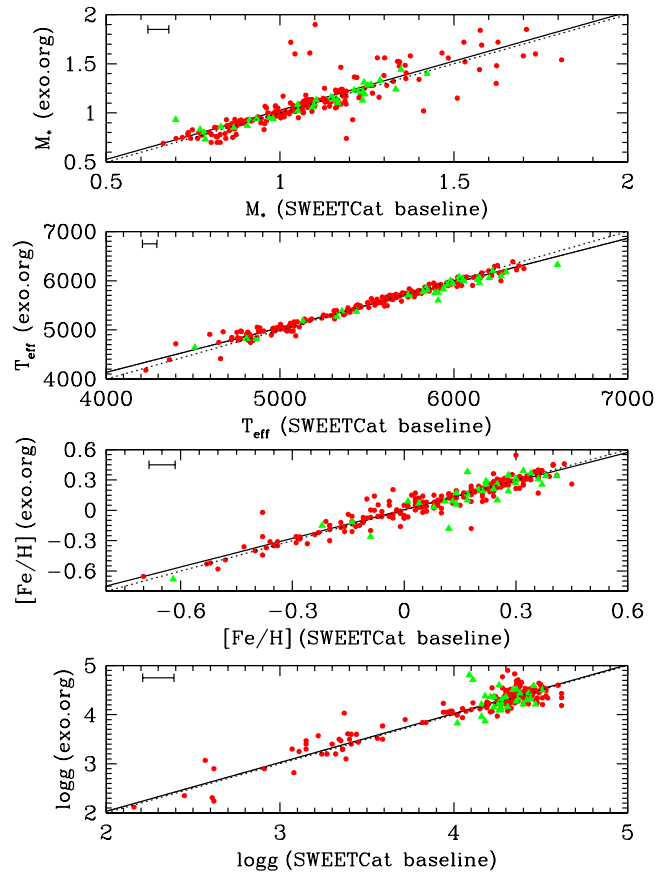}}
\caption{Same as Fig.\,\ref{fig:compeu} but for the data from exoplanets.org \citep[][]{Wright-2011}.}
\label{fig:comporg}
\end{figure}

\begin{figure}[t!]
\resizebox{\hsize}{!}{\includegraphics[angle=0]{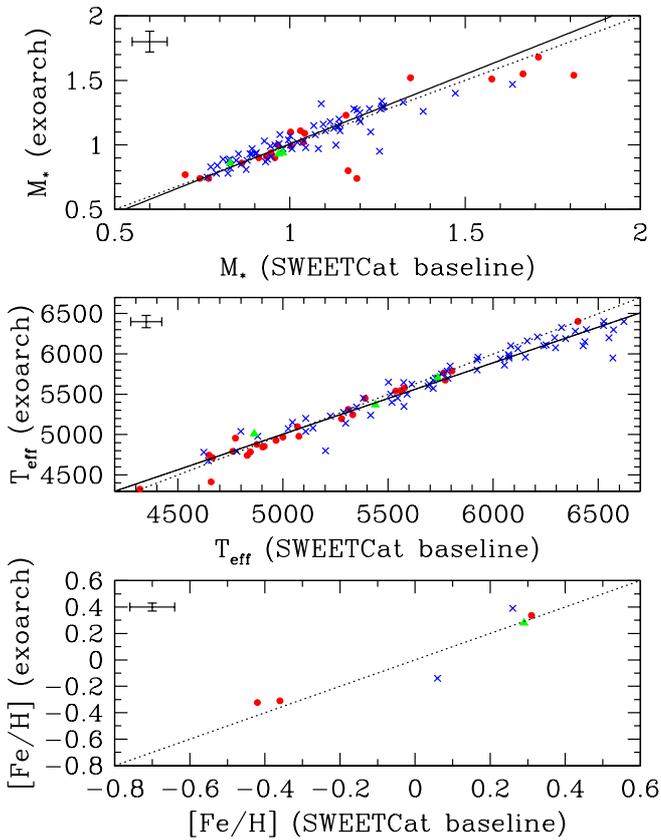}}
\caption{Same as Fig.\,\ref{fig:compeu} but for the data from the NASA Exoplanet Archive. Transiting planets and radial-velocity planets are denoted by crosses and dots, respectively.}
\label{fig:comparch}
\end{figure}

\begin{figure}[t!]
\resizebox{\hsize}{!}{\includegraphics{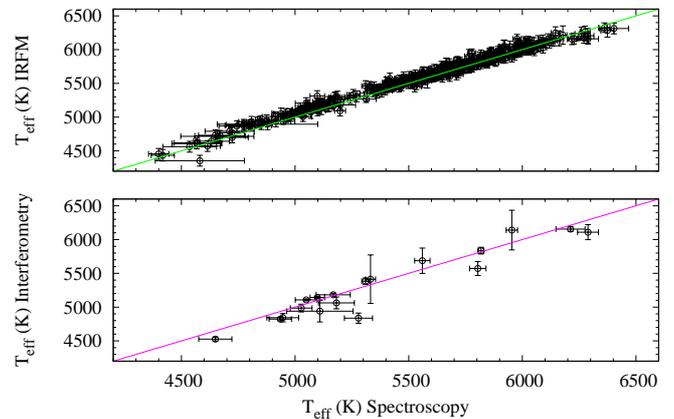}}
\caption{Comparison of the effective temperatures derived using the baseline methodology in the catalogue
with values derived using the IRFM and interferometry. As in \citet[][]{Tsantaki-2013}.}
\label{fig:compteff}
\end{figure}

\begin{figure}[t!]
\resizebox{\hsize}{!}{\includegraphics{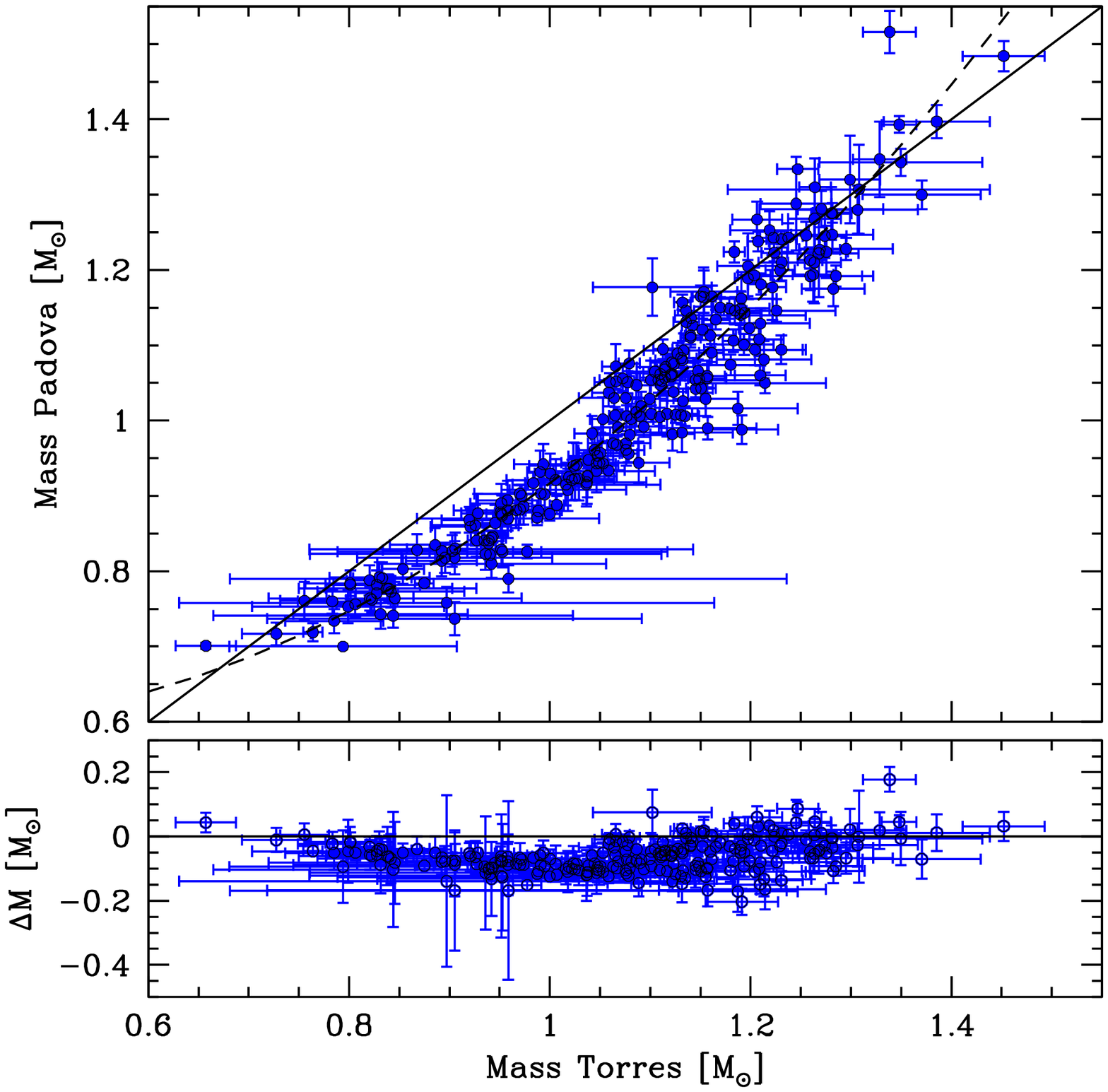}}
\caption{Comparison between the masses for FGK dwarfs derived using the Padova isochones and the \citet[][]{Torres-2010} calibration. The solid line
represents the 1:1 relation and the dotted line a quadratic fit.}
\label{fig:masses}
\end{figure}

When using ``other'' literature data, care was taken to critically compile what we considered
to be the best values, {i.e. the ones that seem to give the best guarantee of uniformity with respect to our baseline parameters}. 
Individual references are shown in the online catalogue.
We are not, however, in position to guarantee the uniformity
of these results {with respect to our baseline parameters, even if in many cases the parameters compare well with
ours for stars we have in common (see below).} 
%We have thus added one flag to tell whether the set of atmospheric
%parameters was taken from uniform samples (meaning an analysis based on
%high resolution spectra done by our team) or not. 

{Again, it is important to add a word of caution. We are not defending here that parameters derived by
other authors are not ``uniform'' (between themselves) or that they are not of ``high quality''. We only say that 
their consistency with respect to our baseline values is more difficult to assure.}

We decided not to include in the table an estimate for the stellar radius.
First, because this can be computed directly from the other fields \citep[see e.g.][]{Santos-2004a}.
Secondly, because for transit host stars, uniform values for this quantity have
been compiled by other groups in other catalogs \citep[e.g.][]{Southworth-2012}.

In the next sections we will describe the baseline methods used to derive what
we will call baseline stellar parameters{, including the values for the 48 stars presented above}. {We also present comparisons of our baseline parameters
with those presented in other catalogs and the literature. These comparisons provide a reference
for the typical systematic errors existing in the stellar parameters for stars with planets
derived by different teams using different, or even sometimes similar methodologies.}

\subsubsection{FGK stars from radial velocity surveys}
\label{sec:FGK_RV}

The most productive radial velocity planet search surveys concentrated their efforts on the search
for planets around solar-type, FGK dwarfs or sub-giants \citep[for some examples see e.g.][]{Udry-2000,Mayor-2003b,Marcy-2005,Johnson-2007b}.
Further to this, and due to astrophysical constraints imposed by active young stars, most of these targets are old, 
slow rotators \citep[][]{Saar-1997,Santos-2000a,Paulson-2002}, with thousands of well defined weak metallic lines
in their spectra. This makes them ideal targets for a standard spectroscopic analysis using iron line equivalent
widths and ionization and excitation equilibrium principles \citep[we point to][for details on the methodology]{Santos-2004b,Sousa-2008,Tsantaki-2013}.

For more than 10 years our team has been obtaining and compiling high resolution spectra to derive 
uniformly stellar parameters and chemical abundances for stars with planetary mass companions discovered 
by radial velocity surveys \citep[e.g.][]{Santos-2001,Santos-2004b,Santos-2005a,Sousa-2008,Sousa-2011,Sousa-2011b,Tsantaki-2013}. 
This lead us to use our own parameters to establish the baseline for the whole catalogue. Note that in several cases, the parameters derived by our team have not been published in "dedicated" papers, but have rather been included in the discovery papers \citep[for recent examples see][]{Boisse-2012,Marmier-2012}.

The choice of this baseline methodology for the derivation of stellar parameters is anchored on the
extremely good agreement with the values found by methods that are usually considered to
be ``standard''. For instance, the temperatures are in very good agreement with those derived using the
Infra-Red Flux Method \citep[IRFM -- see e.g.][and references therein]{Blackwell-1977,Casagrande-2010} and interferometry, both at the 
low \citep[][]{Tsantaki-2013} and high temperature \citep[][]{Sousa-2008} regimes. This result
can be seen in Fig.\,\ref{fig:compteff}, where baseline temperature values are compared
with those derived using the IRFM and interferometry. The differences between the different methods are very small,
with an offset of $-$32\ and 34\,K, for the comparison with the IRFM and interferometry results, respectively (differences are in the sense ``other''$-$``ours''). 
These offsets are mostly independent of the temperature, and cannot be directly attributed to any of the methods used. For more details see \citet[][]{Tsantaki-2013} and references therein.

\begin{table*}
\begin{center}
\caption{Coefficients, residual standard deviation, and number of stars used for the linear regressions of the form x$_{other}$ = a x$_{this paper}$ + b
presented in the plots of Figs.\ref{fig:compeu}, \ref{fig:comporg}, and \ref{fig:comptep}.  }
\label{tab:fits}
\begin{tabular}{cccccc}
\hline
\hline  
 & Quantity & a & b & RMS & N\\
\hline 
\multicolumn{6}{c}{exoplanet.eu} \\
 & {[}Fe/H{]} & 0.908$\pm$0.023 & 0.056$\pm$0.005 & 0.085 & 277\\
 & \emph{$T{}_{\mathrm{eff}}$} & 0.896$\pm$0.015 & 544$\pm$85 & 127 & 268\\
 & M$_{*}$ & 0.997$\pm$0.064 & 0.040$\pm$0.074 & 0.308 & 278\\
\hline 
\multicolumn{6}{c}{exoplanets.org}\\
 & [Fe/H] & 0.943$\pm$0.021 & 0.006$\pm$0.004 & 0.072 & 245\\
 & \emph{$T{}_{\mathrm{eff}}$} & 0.909$\pm$0.010 & 498$\pm$54 & 72 & 240\\
 & $\log{g}$ & 0.995$\pm$0.002 & 0.042$\pm$0.103 & 0.18 & 238\\
 & M$_{*}$ & 1.000$\pm$0.056 & 0.026$\pm$0.062 & 0.22 & 245\\
\hline 
\multicolumn{6}{c}{NASA Exoplanet Archive}\\
 & \emph{$T{}_{\mathrm{eff}}$} & 0.884$\pm$0.018 & 583$\pm$103 & 109 & 97\\
 & M$_{*}$                                     & 1.073$\pm$0.041 & $-$0.067$\pm$0.039 & 0.14 & 96\\
\hline 
\multicolumn{6}{c}{TEPCat}\\
 & {[}Fe/H{]} & 1.033$\pm$0.130 & $-$0.053$\pm$0.027 & 0.13 & 39\\
 & \emph{$T{}_{\mathrm{eff}}$} & 0.845$\pm$0.029 & 852$\pm$168 & 106 & 39\\
 & M$_{*}$ & 0.858$\pm$0.062 & 0.154$\pm$0.068 & 0.08 & 39\\
\hline 
\end{tabular}
\end{center}
\noindent
%Notes: Columns 3 (a) and 4 (b) represent the robust linear fits of x$_{other}$ = a x$_{this paper}$ + b. 
%Column 5 represents the residual standard deviation of the regression lines. 
\end{table*}

To keep uniformity, for all FGK dwarfs with {baseline atmospheric parameters}, stellar masses have been derived using a
uniform method. For simplicity, we computed them with the calibration of \citet[][]{Torres-2010}, using as input our spectroscopic parameters. A small correction was however applied, as follows. 
The values derived using this calibration are in general similar to the ones obtained using the web interface based on Padova 
isochrones \citep[][]{LicioDaSilva-2006}\footnote{http://stev.oapd.inaf.it/cgi-bin/param} -- Fig.\,\ref{fig:masses}. However, a general offset is present that is a 
function of stellar mass. This offset was already discussed in \citet[][]{Torres-2010}. In order to correct for this offset, we fitted a quadractic function
to the plot in Fig.\,\ref{fig:masses}:
\begin{equation}
\mathrm{M_{iso} = 0.791 \times M_{T}^2 - 0.575 \times M_{T} + 0.701}
\end{equation}
where M$_\mathrm{iso}$  and M$_\mathrm{T}$ denote the stellar masses derived using the Padova isochrones and the
Torres et al. calibration, respectively. This equation was used to correct for the mass values listed in the catalogue.

{Errors in the stellar mass were also computed using the ``corrected'' Torres et al. calibration. The values were derived by means of a
Monte Carlo analysis, where in each case 10\,000 random values of effective temperature, surface gravity, and stellar metallicity were 
 drawn assuming a gaussian distribution from the derived uncertainties. The resulting mass distribution is
used to derive the central value (the mass) and the 1-sigma uncertainty. The intrinsic error in the Torres et al. calibration was
also quadratically added to the final uncertainly.}

In Figs.\,\ref{fig:compeu}, \ref{fig:comporg}, and \ref{fig:comparch} we compare our {baseline parameters} for stars detected in the context
of radial velocity surveys with those listed in the Extrasolar Planets Encyclopaedia \citep[][]{Schneider-2011}, exoplanets.org \citep[][]{Wright-2011}, and the NASA 
Exoplanet Archive\footnote{http://exoplanetarchive.ipac.caltech.edu/}. {As mentioned above,} green triangles denote the parameters for the 48 stars presented in this paper. The general trends show a good agreement, though some systematic effects are present. In Table\,\ref{tab:fits} we present the 
coefficients of the linear fits to the data. These may be used to correct for the systematic trends. Due to the small number of points available, no 
fit was done for the comparison of metallicities with the data from the NASA 
Exoplanet Archive (this archive only has metallicities for a minority of the stars listed).
Note also that several important outliers appear in the plots. This
shows the need for a careful and uniform derivation of stellar parameters in any case-by-case analysis of stars with planets.
Finally, note that in several cases the parameters listed in the former two catalogs mentioned above were taken from our
own sources, a fact that contributes to the improvement of the agreement seen in the plots.

As mentioned above, {for FGK dwarfs which do not have ``baseline" spectroscopic parameters}, stellar parameters
were compiled from the literature. Whenever possible, we used sources for which the stellar parameters
compare well with our own values \citep[e.g., the SPOCS catalogue][]{Valenti-2005} -- see \citet[][]{Sousa-2008} for a comparison.

\subsubsection{FGK stars with transiting planets}

For all FGK stars with transiting planets for which we could obtain a high resolution spectrum,
{atmospheric parameters} and masses were derived using the same methodology described
in the previous section. As before, most of these parameters have already been 
published in dedicated papers \citep[e.g.][-- see also Mortier et al. 2013, in prep.]{Santos-2006a,Ammler-2009} or
in planet discovery papers where the spectroscopic analysis was done
by our team \citep[see][for a recent example]{Santerne-2012}. This guarantees the {best possible uniformity} of the results.

For stars with transiting planets, surface gravities were also derived using the information
coming from the transit light curves. Indeed, surface gravities are typically very difficult to determine accurately
through spectroscopy. For stars with a transiting planet, however, the
surface gravity can be determined more directly. Purely from transit
photometry, the stellar density can be calculated from \citet[][]{Seager-2003}:
\begin{equation}
\rho_{\ast} + k^3\rho_p = \frac{3\pi}{GP^2}\left(\frac{a}{R_{\ast}}\right)^3
\end{equation}
Since the constant coefficient $k$ is usually small, the second term on the left is negligible.
All parameters on the right come directly from the transit light curve (in the present paper
these were taken directly from transit analysis papers in the literature).
With this stellar density, combined with the effective temperature and 
metallicity from the spectroscopic analysis, the surface gravity can be determined
through isochrone fitting \citep[see e.g.][]{Sozzetti-2007}. As presented in Mortier et al. (2013, in prep.),
for this step, we used the PARSEC isochrones \citep[][]{Bressan-2012} and a $\chi^2$ minimization process for the fitting.

The temperatures and metallicities derived using the ionization and excitation equilibrium of iron lines
have shown to be mostly independent of the adopted surface gravity \citep[][]{Torres-2012}. This is due
to the relatively low sensitivity of \ion{Fe}{i} lines (used to constrain the temperature and metallicity) to
changes in $\log{g}$. For example, if we derive the effective temperature and metallicity for the Sun
using the adopted methodology and line-lists but fixing $\log{g}$ to 3.0 (a strong $\sim$1.5\,dex difference), the
derived effective temperature and metallicity values are only $\sim$250\,K and $\sim$0.10\,dex higher,
respectively, than the adopted solar values. As such, the temperatures and metallicities derived with our adopted spectroscopic method 
can be used as reference values even if the derived spectroscopic surface gravities differ from
those derived using the transit light curve \citep[and the stellar density --][]{Sozzetti-2007}. A more detailed
discussion about this issue will be presented in Mortier et al. (in prep.).

As mentioned above, the effective temperatures derived by the adopted methodology are in very
good agreement with those derived by the IRFM. This implies that the stellar radii that we can
derive using these parameters are probably as accurate as one can guarantee.

\begin{figure}[t!]
\resizebox{\hsize}{!}{\includegraphics[angle=0]{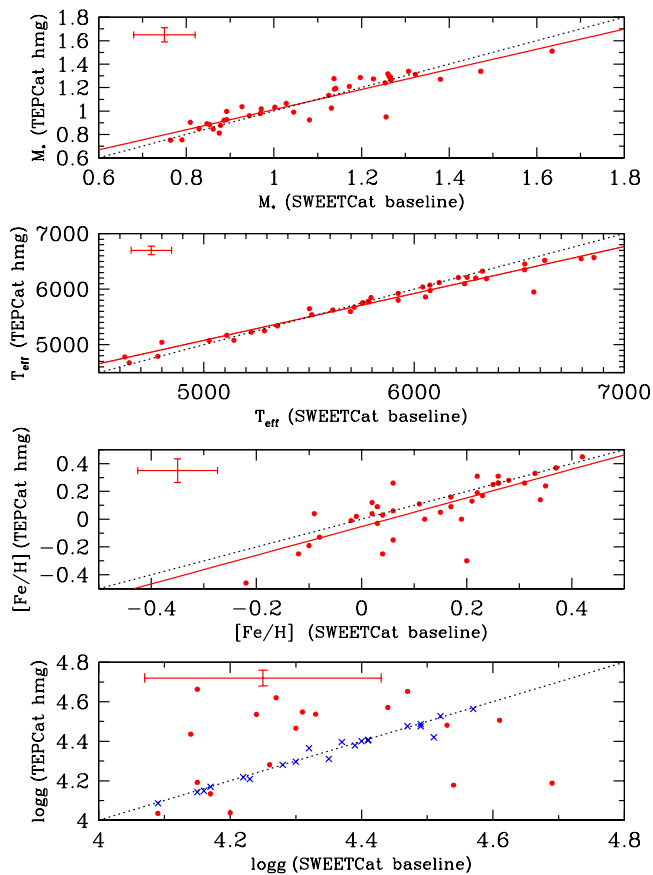}}
\caption{Same as Fig.\,\ref{fig:compeu} but for the data from the ``homogeneous'' part of the TEPCat catalogue \citep[][]{Southworth-2012}. {In the lower panel, crosses denote surface gravities derived using the transit light curve.}} 
\label{fig:comptep}
\end{figure}

In Fig.\,\ref{fig:comptep} we compare our {baseline parameters} for FGK stars with transiting planets
with those presented in the ``homogeneous table'' of the TEPCat catalogue \citep[][]{Southworth-2012}\footnote{http://www.astro.keele.ac.uk/jkt/tepcat}. On the $\log{g}$ plot (lower right),
crosses denote a comparison with our surface gravities derived using the transit light curve, while dots denote a comparison with our purely spectroscopic values. X-axis error bars
refer to the typical spectroscopic uncertainties. Due to the very good agreement, we decided not to present any fit for the $\log{g}$ comparison. For all parameters compared, the results show again a good agreement. There is, however, a small offset on the metallicities between the two samples, and perhaps more important, a general trend on the temperature scales. 
This temperature scale difference may conduct to the derivation of significant different values for the planetary radii (in particular for the higher temperature stars).
The dispersion in the $\log{g}$ comparison denoted the higher errors present in the pure spectroscopic analysis.

Whenever we did not have access to a high resolution spectrum (mostly for the cases of planets detected as
part of the Kepler and WASP surveys),
priority was given to studies and compilations such as the TEPCat catalogue \citep[][for transiting planets]{Southworth-2012}.
For the remaining stars, planet discovery papers were often used as the source for the stellar parameters. In some cases,
the methodologies used are similar to the ones adopted for the majority of the stars in our catalogue.

\subsubsection{Giant and evolved stars}

The determination of stellar parameters for cool, giant stars is a matter of strong debate in the literature, with
several authors raising doubts about the zero point of the metallicity scale in these objects 
\citep[e.g.][]{Taylor-2005,Cohen-2008,Santos-2009,Santos-2012}. Although the exact reasons are still
not clear, these problems may have even lead to a significant discrepancy in studies done by different authors concerning the metallicity-giant
planet correlation in giants \citep[see debate in ][]{Pasquini-2007,Hekker-2007,Ghezzi-2010}.

To guarantee the maximum homogeneity degree in the parameter scale used in the present paper, we decided
to adopt as baseline the recent study by Mortier et al. (2013, in prep.) %\citet[][]{Mortier-2013c}, 
where the parameters for 71 evolved stars
with planets were derived using the same iron line ionization and excitation equilibrium method used for
the study of FGK dwarfs. 
%In this case, a new line-list specially tested for cool stars \citep[see][]{Tsantaki-2013} was used.
For the remaining stars, values were compiled from the literature, both from the discovery papers or from
other compilations/catalogs \citep[e.g.][]{Luck-2007,Soubiran-2010}.

Finally, since the mass calibration presented in \citet[][]{Torres-2010} is not valid for giant stars, the masses for all stars with $\log{g}$
values lower than $\sim$4.0 were derived using the Padova isochrones \citep[][]{LicioDaSilva-2006}. 

\subsubsection{M-dwarfs}

The derivation of M-dwarf atmospheric parameters is a challenging task.
Due to the difficulty in deriving precise values for the effective temperature and metallicity based on spectral fitting 
procedures \citep[e.g.][]{Valenti-1998,Woolf-2005,Bean-2006,Onehag-2012}, most determinations of their values are based
on calibrations using colors \citep[][]{Bonfils-2005,Johnson-2009,Casagrande-2008,Schlaufman-2010,Neves-2012} or spectroscopic 
indices \citep[e.g.][]{Terrien-2012,Rojas-Ayala-2012,Mann-2012, Neves-2013}.

For consistency reasons, in this paper we used the photometric calibration of \citet[][]{Neves-2012} as out baseline to measure the metallicity. In the case where HARPS spectra were available, however, the parameters were derived using the new \citet[][]{Neves-2013} spectroscopic calibration. Both \citet[][]{Neves-2012} and  \citet[][]{Neves-2013} calibrations use the same metallicity scale,
assuring thus uniformity in the results. The [Fe/H] uncertainties of the two calibrations are assumed to be 0.20 and 0.10 dex, respectively.
The metallicity scale used compares very well with other estimates from the literature \citep[see e.g.][]{Neves-2012}.

%In this paper, for consistency reasons we decided to adopt as baseline the metallicities derived and presented in \citet[][]{Neves-2012} for 19 M-dwarfs,
%exception made, for the Kepler stars and those for which we could compile HARPS spectra (8 stars).
%In the latter case, the parameters were derived using the new \citet[][]{Neves-2013} standard. Both methods use the same metallicity scale, assuring thus uniformity in the results. 
%The [Fe/H] uncertainties are assumed to be 0.20 and 0.10 dex, estimated from the dispersion of the calibrations of \citet[][]{Neves-2012} and \citet[][]{Neves-2013}, respectively.

\begin{figure*}[t!]
\includegraphics[width=9cm]{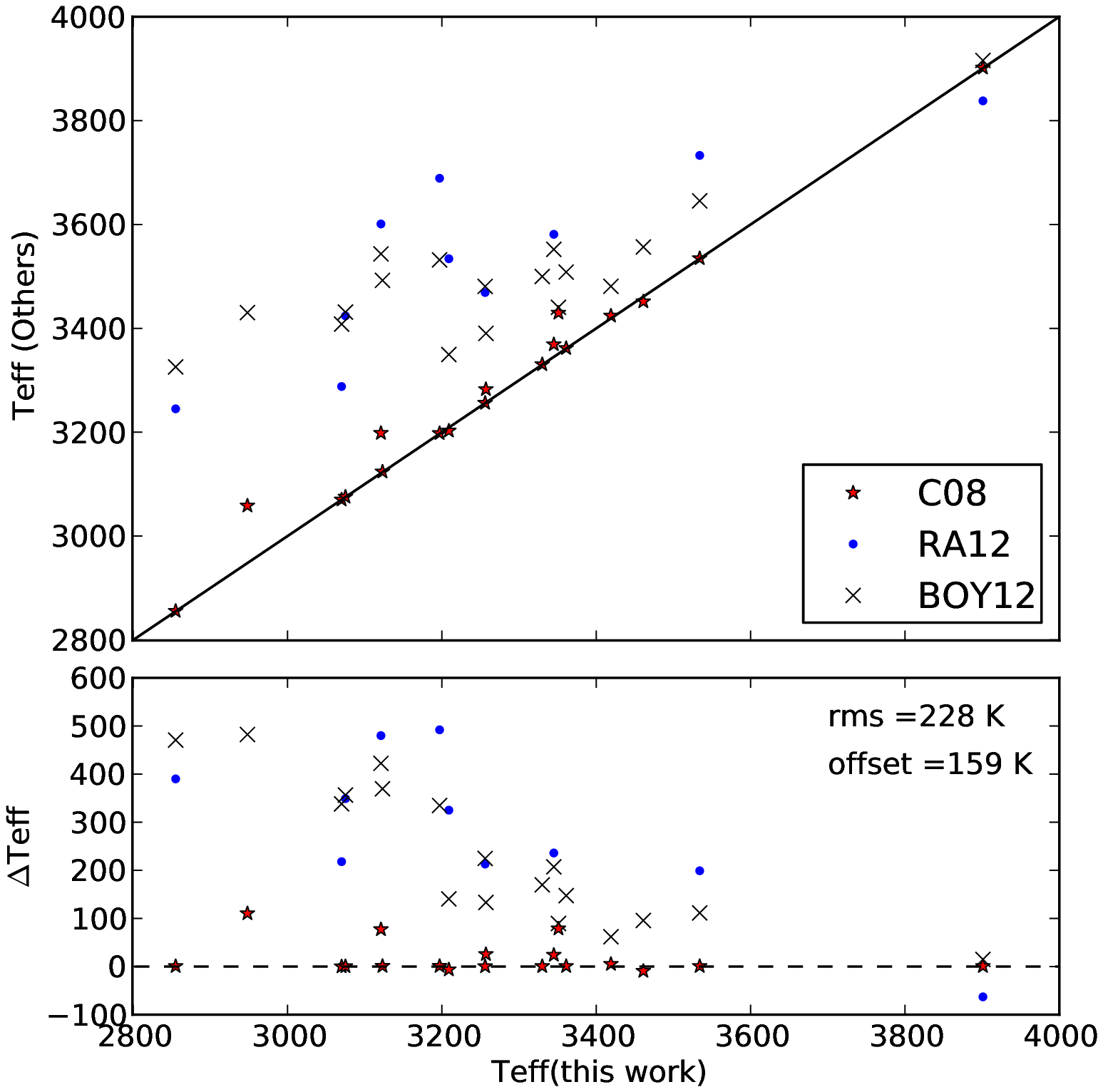}
\includegraphics[width=9cm]{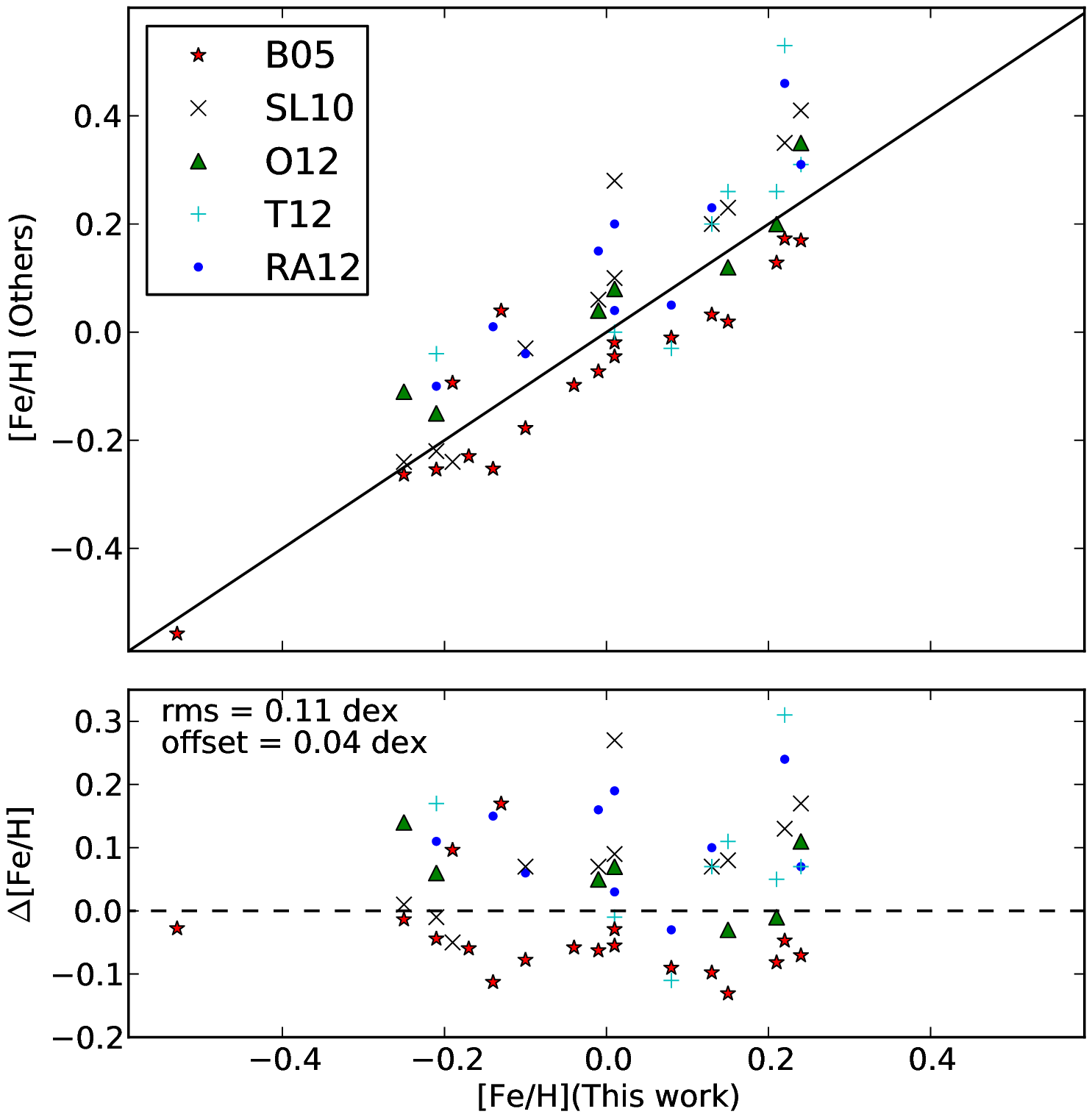}
\caption{Comparison of the effective temperatures (left) and metallicities (right) for M-dwarfs listed in this catalog with values obtained in the literature for using other calibrations. See text for more details.} 
\label{fig:Mdwarfs}
\end{figure*}

\begin{table}
\begin{center}
\caption{Mean offsets and number of stars used to compare the metallicity and temperature scales for M-dwarfs.}
\label{tab:Mdwarfs}
\begin{tabular}{lcc}
\hline
\hline  
Comparison study & <offset> & N\\
\hline 
\multicolumn{3}{c}{[Fe/H]}\\
All stars & 0.04 & -- \\
B05 & $-$0.04 & 18\\
SL10 & 0.08 & 11\\
O12 & 0.06 & 7\\
T12 & 0.08 & 8\\
RA12 & 0.11 & 10\\
\hline 
\multicolumn{3}{c}{T$_\mathrm{eff}$} \\
All stars & 159 & -- \\
C08 & 37 & 18  \\
RA12 & 283 & 10\\
BOY12 & 231 & 18\\
\hline 
\end{tabular}
\end{center}
\noindent
%Notes: Columns 3 (a) and 4 (b) represent the robust linear fits of x$_{other}$ = a x$_{this paper}$ + b. 
%Column 5 represents the residual standard deviation of the regression lines. 
\end{table}

Effective temperatures for all the stars in this paper, except for the case of the Kepler stars (see below), were derived using the calibrations based on the V$-$J, V$-$H, and V$-$K colors
presented in \citet[][]{Casagrande-2008}. These are based on the MOITE method which is an optical extension of the IRFM \citep[][]{Blackwell-1977}.  For the cases where HARPS spectra were available, the spectroscopic calibration of \citet[][]{Neves-2013} was used instead. 
This calibration used the Casagrande parameters as baseline, meaning that all values are on the same scale and have the 
same accuracy. The uncertainty in T$_\mathrm{eff}$ for the \citet[][]{Casagrande-2008} was computed by adding the propagation of the errors of the V and infrared 
photometry \citep{Skrutskie-2006} taken to calculate the calibrations with the estimated error of the calibration (150\,K).
We assume an error of 150K for the  \citet[][]{Neves-2013} relation.

The stellar masses were derived using the K-band empirical calibration of \citet[][]{Delfosse-2000}. Mass uncertainties are estimated to be 10\%.
The surface gravities were derived using Newton's law from the mass and the radius derived using the empirical relations of \citet{Boyajian-2012}. We estimate a 10\% uncertainty for the radii measurements. The uncertainties of the surface gravity are calculated by propagating the errors of the mass and radius. 
As for the remaining stars, parallaxes were taken from Simbad, except when otherwise mentioned.% for GJ1214 where the photometric relation of \citet{Anglada-Escude-2012} was used.

Given the differences in the methodologies used to derive stellar parameters for FGK stars (see above) and those used here for M-dwarfs, for M-dwarfs we cannot guarantee that
the parameters derived are on the same scale as those derived for the FGK dwarfs. However, our choice gives us some confidence that
the values for their parameters are homogeneous between themselves.

In Fig.\,\ref{fig:Mdwarfs} we compare the metallicity and effective temperature values derived using the methodology described above
with those presented by other authors or derived using other calibrations. As denoted in the insets, different symbols denote different sources:
\citet[][C08]{Casagrande-2008}, \citet[][RA12]{Rojas-Ayala-2012}, \citet[][BOY12]{Boyajian-2012} concerning the effective temperatures,
and \citet[][B05]{Bonfils-2005}, \citet[][SL10]{Schlaufman-2010}, \citet[][O12]{Onehag-2012}, \citet[][T12]{Terrien-2012}, and \citet[][RA12]{Rojas-Ayala-2012}
concerning metallicities. The results show that in general, and on average, the values used in this catalog are reasonably well correlated with those derived in the literature (or derived using 
specific calibrations). The major difference concerns the effective temperatures, for which our values agree very well with the ones
derived using the \citet[][]{Casagrande-2008} IRFM calibration, but present a significant offset with respect to other literature values, specially for the lower temperature stars. The
agreement with the Casagrande et al. determinations come with no surprise, since our temperature scale was calibrated using their
values as reference. In Table\,\ref{tab:Mdwarfs} we list the average offsets between the different sets of data as well as the number of stars used
for the comparison shown in Fig.\,\ref{fig:Mdwarfs}. All values denote the differences in the sence ``literature'' $-$ ``this work''.

For Kepler M-stars, due to the difficulty in gathering either high resolution spectra or reliable photometry, 
we opted to take the parameters from the TEPCAT catalogue \citep[][]{Southworth-2012}, directly from the discovery papers,
or from updated papers from the Kepler team. 

%The parallaxes and its uncertainties are mostly taken from Hipparcos \citep[][]{vanleewen-2007}, but also from van Altena (1995) (GJ1214), Riedel (2010) (GJ3634), and Anglada-Escude (2012) (GJ317).
%In order to calculate the stellar mass, the Ks photometric measurements from 2MASS \citep[][]{Cutri-2003} were transformed into CIT photometry \citep[][]{Frogel-1978,Elias-1982} using the \citet[][]{Carpenter-2001}. The parallax is then used with the infrared Ks photometry from 2MASS to calculate absolute K-band Magnitudes which are input into the M/L K-band relationship of \citet[][]{Delfosse-2000}.

\subsubsection{General comments and the online catalogue}

\begin{figure}[t!]
\resizebox{\hsize}{!}{\includegraphics[angle=270]{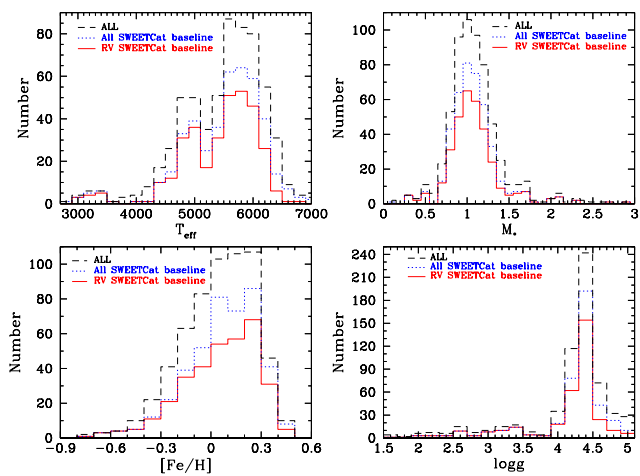}}
\caption{Histograms with the distributions of different stellar parameters in our catalogue.}
\label{fig:histo}
\end{figure}

In Fig.\,\ref{fig:histo} we plot the distribution of effective temperatures, metallicities, surface gravities, and masses that
are listed in our catalogue. Besides the whole distribution, we also plot the histogram for the sample of FGK stars with {derived baseline}
stellar parameters, as well as the subsample of FGK stars with planets discovered using the radial velocity method. 

%In Fig.\,\ref{fig:hr} we further plot a modified HR diagram, where the luminosity was replaced by the surface gravity.  Only stars whose parameters
%were derived with our ``homogeneous" analysis are presented. For comparison, different symbols are used for transit host stars.
%As expected, both plots show that most of the stars with planets have parameters typical of FGK dwarfs.

%\begin{figure}[t!]
%\resizebox{\hsize}{!}{\includegraphics[angle=270]{teff_logg_HR.ps}}
%\caption{Effective temperature vs. $\log{g}$ plot for all the stars in our catalogue with ``homogeneous'' derived parameters. M-dwarfs were not included.}
%\label{fig:hr}
%\end{figure}

The complete table with compiled stellar parameters for planet host stars is available online at \url{https://www.astro.up.pt/resources/sweet-cat}. Besides the html version, the reader can download an ascii file with all the fields. Improvements on this online table will be done
on a continuous basis.

\section{Conclusions}
\label{sec:conclusions}

{In this paper we present new spectroscopic atmospheric parameters and masses for a sample of 48 stars with planets
discovered in the context of different radial velocity planet search programs.

These parameters are then included in
a new catalogue of stellar parameters for FGK and M stars with planets. 
The stellar parameters in this catalog are compiled from literature sources in a way that optimizes the uniformity of the values, making them
more suitable for statistical studies of stars with planets. 
The catalogue will be updated as new planet hosts appear in the literature. We will also continue
our effort to determine on a regular basis uniform stellar parameters from high resolution and high
S/N spectra. New parameter values may be added to the catalog even before a paper is published to present them.
}

At the time this paper is being published, the parameters listed in the catalogue come from literature sources,
both published or to be published soon. Without all these studies the present compilation would not
have been possible. Although we do not encourage, we understand that for simplicity
the user may wish to cite only the present paper if using the catalogue in a statistical way. We strongly suggest, however, that in studies 
of individual stars the original source of the parameters is also cited.

In its present form, the catalogue presents, besides basic parameters, a compilation of atmospheric
parameters and masses for all planet host stars known. In the future the catalogue may be expanded
to add additional stellar parameters of interest, such as the projected rotational velocity ($v\,\sin{i}$),
the rotational period, and the chromospheric activity level ($\log{R'_{HK}}$). Furthermore, we
are considering to compile also chemical abundances for elements other than iron as long
as uniform sources exist \citep[e.g.][]{Adibekyan-2012b}. 

\begin{acknowledgements}
We would like to thank Paulo Peixoto for the rapid construction of the web page with the online catalogue.
This work has made use of the Simbad database.
 This work was supported by the European Research Council/European Community under the FP7 through Starting Grant agreement 
 number 239953 and by Funda\c{c}\~ao para a Ci\^encia e a Tecnologia (FCT) in the form of grant reference PTDC/CTE-AST/098528/2008. 
 EDM, SGS, VN, and VZhA also acknowledge the support from FCT in the form of fellowships reference SFRH/BPD/76606/2011, 
 SFRH/BPD/47611/2008, SFRH/BD/60688/2009, and SFRH/BPD/70574/2010. GI acknowledges financial support from the Spanish Ministry
project MICINN AYA2011-29060.
\end{acknowledgements}

\bibliographystyle{aa}
\bibliography{santos_bibliography}

\end{document}